\documentclass[fleqn,preprint,5p]{elsarticle}

\usepackage{hyperref}
\usepackage{color}
\usepackage{amsmath, amssymb, amsfonts}
\usepackage{graphicx}
\usepackage{hyperref}

\journal{Physics Letters B}

\def\beq{\begin{equation}}
\def\eeq{\end{equation}}
\def\beqa{\begin{eqnarray}}
\def\eeqa{\end{eqnarray}}
\def\ben{\begin{enumerate}}
\def\een{\end{enumerate}}
\def\bit{\begin{itemize}}
\def\eit{\end{itemize}}

\begin{document}

\begin{frontmatter}

\title{%Right-handed currents and experimental data analysis\\
%through heavy neutrinos' mass spectra, CP-phases and mixing parameters, \\as exemplified by %the $pp\to lljj$ CMS excess data
Heavy neutrinos and 
%experimental data analysis\\
%through heavy neutrinos' mass spectra, CP-phases and mixing parameters, \\
%as exemplified by 
%the excess in 
the $pp\to lljj$ CMS data\tnoteref{t1}}

%\tnotetext[mytitlenote]{Fully documented templates are available in the elsarticle package on \href{http://www.ctan.org/tex-archive/%macros/latex/contrib/elsarticle}{CTAN}.}

\tnotetext[t1]{This article is registered under preprint number: 1504.05568 [hep-ph]}

\author{Janusz Gluza}
%\ead{janusz.gluza@us.edu.pl}
\author{Tomasz Jeli\'nski\corref{ca1}}
\cortext[ca1]{Corresponding author}
\ead{tomasz.jelinski@us.edu.pl}
\address{Institute of Physics, University of Silesia, Uniwersytecka 4, 40-007 Katowice, Poland}

\begin{abstract}
We show that the excess in the  
$pp \to ee jj$ CMS data can be naturally interpreted within the
Minimal Left-Right Symmetric model (MLRSM), keeping $g_L = g_R$, if CP phases and non-degenerate masses of heavy neutrinos are taken into account.  As an additional 
benefit, a natural interpretation of the reported ratio (14:1) of the opposite-sign (OS)
$pp\to l^\pm l^\mp jj$ to the same-sign (SS) $pp\to l^\pm l^\pm jj$ lepton signals is possible. Finally, a suppression of muon pairs with respect to electron pairs in the $pp \to lljj$ data is obtained, in accordance with experimental data. If the  excess in the CMS data survives in the future, it would be a first clear hint towards presence of heavy neutrinos in right-handed charged currents with specific CP phases, mixing angles and masses,  which will have far reaching consequences for particle physics directions.
\end{abstract}


\end{frontmatter}

%\linenumbers

\section{Introduction}

\begin{figure}[h!]
\begin{center}
\includegraphics[scale=0.5]{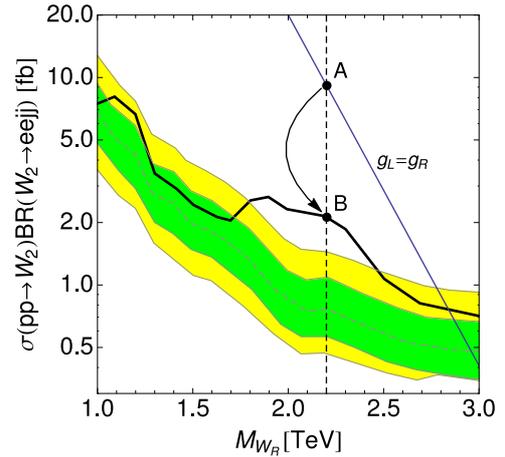}
\caption{CMS data for production of the first generation leptons with two jets in $pp$ collision with $\sqrt{s}=8\,\mathrm{TeV}$ \cite{Khachatryan:2014dka}.  Blue solid line shows CMS estimation of the cross section in the  MLRSM model with $g_L=g_R$, diagonal heavy neutrino couplings and $M_N=M_{W_2}/2$.  Explaining excess in the data around $M_{W_2}\sim2.2\,\mathrm{TeV}$ (point B) requires refinement of those assumptions (point A).
}\label{CMS}
\end{center}
\end{figure}

LHC is a perfect laboratory to test Beyond Standard Model (BSM) scenarios. Recently, the CMS Collaboration announced an interesting 
%yet statistically not significant, 
excess in  data, see point B on Fig. \ref{CMS}. This point is related to the process $pp\to eejj$ collected by $\sqrt{s}=8\,\mathrm{TeV}$ LHC corresponding to an integrated luminosity $19.7$ $\mathrm{fb}^{-1}$ \cite{Khachatryan:2014dka}.
Several analyses \cite{Deppisch:2014qpa, Heikinheimo:2014tba, Deppisch:2014zta,Aguilar-Saavedra:2014ola} showed that this excess can be interpreted as a signal of charged gauge boson $W_2^\pm$ with mass about $2.2\,\mathrm{TeV}$  in the Left-Right symmetric model \cite{Mohapatra:1974gc,Senjanovic:1975rk,Mohapatra:1980yp}. It is possible when gauge couplings connected with left and right $SU(2)$ groups  are not equal to each other. For a case $g_L=g_R$ see point A on Fig.~\ref{CMS} (the measured cross section is suppressed by a factor of $\gamma_{CMS}=0.23$ when compared with scenario in which $g_L=g_R$).  
Moreover,  the number of events with same-sign (SS) leptons to the number of events with opposite-sign (OS) leptons is 
\begin{equation}
r=\frac{ N_{SS}}{ N_{OS}}=\frac{1}{14},\label{r}
\end{equation}
 and, finally, no excess in $\mu\mu$ channel has been reported \cite{Khachatryan:2014dka,Khachatryan:2015gha}.

Theoretical analyses of the left-right symmetric models speeded up considerable in recent years  
\cite{Maiezza:2010ic,Tello:2010am,Nemevsek:2011hz,Nemevsek:2012cd,Das:2012ii,Adelman:2012py,
Nemevsek:2012iq,Dev:2013oxa,Mohapatra:2013cia,Chen:2013foz,
Bertolini:2014sua,Maiezza:2014ala,
Vasquez:2014mxa,Aydemir:2014ama,Staub:2014pca,Dutta:2014dba,
Mahajan:2014nca,deAnda:2014dba,Senjanovic:2014pva,Maiezza:2015lza, Dev:2015vra,Senjanovic:2015yea,Bambhaniya:2013wza}, after the LHC has started its operation. It is not surprising as this collider is operating at   highest available so far energies, 
%comparing to previous machines
which means that new states of matter or new interactions can be probed more effectively.  For instance, the left-right symmetric models offer an elegant, dynamical explanation for suppression of right handed currents at low energies, and it might be that finally LHC can see them directly in experimental data analyses.\footnote{For other relevant arguments in favor of left-right symmetry, see e.g. \cite{Mohapatra:1986uf}.}

 Discovery of right-handed currents and new elementary states of matter in
form of a
charged heavy gauge boson and heavy neutrinos would be of paramount importance for our understanding of Physics in microscale.
It would also impact Physics in macroscale. For instance, details of leptogenesis depend on CP phases of decaying particles, or Big Bang Nucleosynthesis and the dark matter problem raise
questions about the matter content of the Universe \cite{unbound,Giunti:2007ry} (neutrinos may be responsible  for the matter-dominated flat Universe in which we live).

It is natural that experimental data analysis employs simplifications of theoretical models which quite often, thinking in terms of BSM, are much more complicated than worthy Standard Model theory. However, in this way conclusions can be distorted or even some interesting and natural scenarios can be 
%lost. 
overlooked. We think that our discussion here is a good and important example showing that including some additional theoretical issues into analysis can finally pay back in terms of better understanding of experimental results.
 
%It might happen, for instance, when CP parities of heavy neutrinos are neglected. 
%{\bf Such (?)} CP-sensitive processes are e.g. the neutrinoless double beta decay $(\beta %\beta)_{0\nu}$ or the {\bf LHC process (?)} where two leptons and two jets are produced, $pp %\to ll jj$. In these processes, amplitudes containing non-SM corrections, include heavy neutrino mixing matrix elements $K_R$, and taking into account CP parities of heavy neutrinos, some of $K_R$ elements become complex. Then destructive interferences are present in amplitudes, leading to substantially different cross-sections.

In this paper, we show how including details of heavy neutrinos mass spectrum, their CP phases and 
%heavy neutrino 
non-trivial mixing matrix can change a picture, leading to natural interpretation of the data within MLRSM, with $g_L = g_R$ and relatively light $W_2^\pm$ charged gauge boson mass.\footnote{In this work, we do not discuss which scenarios are more natural, those with $g_L=g_R$, or those without it. If we demand strict unification of gauge couplings then scenarios with $g_L \neq g_R$ seem to be more appropriate \cite{Deppisch:2014qpa,Heikinheimo:2014tba,Deppisch:2014zta}. However, also such models need additional modifications, like new intermediate scales or symmetries. This problem  is not of the main importance for our work, and we shall remain with the simplest possibility, which is $g_L=g_R$. MLRSM has an additional advantage as $g_L=g_R$ preserves $P$ symmetry and simplifies model parameters, e.g. gauge bosons mass relations.} In other words, we can get down from  point A to point B on Fig. \ref{CMS}, while holding $g_L=g_R$. We can also accommodate value of $r$ in (\ref{r}) and explain a shortage of muon pairs.  
%An example is specific, but conclusions which we can derive are more general.
% Namely, considering MLRSM example, we want to point out that 
%details on interference effects connected with splitting of heavy neutrino masses, non-%diagonal couplings 
%of heavy neutrinos to $W_2$ and charged leptons or CP phases of heavy neutrinos, all of them are important for interpreting experimental data, 
%An example is specific, but conclusions which we can derive are more general
%as heavy neutrinos are present within many BSM models. 

We shall consider production of $W_2$ which further decays to a charged lepton $l_i$ ($i=1,2,3$) and an on-shell heavy neutrino $N_a$ ($a=1,2,3$) \cite{Keung:1983uu}. The latter further decays mainly  via 3-body process $N_a\to l_jjj$ leading to two jets and two charged leptons in the final state:
\beqa\label{pp_lljj}
pp\to W_2^{\pm}\to l_i^{\pm} N_a\to l_i^{\pm} l_j^{\mp}jj,&&\quad\textrm{(OS)}\\
pp\to W_2^{\pm}\to l_i^{\pm} N_a\to l_i^{\pm} l_j^{\pm}jj.&&\quad\textrm{(SS)}
\eeqa
%We analyze these processes and point out setups, other than departing from $g_L=g_R$, which could be the origin of the above-mentioned excess.  

We have left considering distributions of kinematical variables of leptons and jets as well as cuts issue in the discussed processes for future studies. 
%When larger amount of data is available such analysis could be in order.

\section{Heavy neutrino interactions  and their CP parities}
We work within the MLRSM model in which $g_L=g_R$, $v_L=0$ and $\kappa_2=0$ (what results in no $W_L-W_R$ mixing). 
%parameter $\xi=0$). 
%
The scale of breaking $SU(2)_R$ is set to 
$v_R=4.77\,\mathrm{TeV}$,
%$v_R=5.2\,\mathrm{TeV}$, 
%
such that the mass of $W_2$ is about $2.2\,\mathrm{TeV}$ (see Fig.~\ref{CMS}).
%such that the mass of $W_2$ is about $2.4$ TeV. 
%{\bf Setting $M_N=M_{W_2}/2=1.1\,\mathrm{TeV}$ could be also in order to directly address CMS analysis.} 
%
Moreover, to simplify our considerations, let us assume that the scalar potential parameters are chosen such that all scalar particles beside the lightest Higgs boson have masses of order $v_R$. We leave discussion of their influence on $pp\to lljj$ for future studies. 

Neutrino mass matrix is chosen to be of the form
\begin{equation}
M_{\nu}=\left(
\begin{array}{cc}
0 & M_D\\
M_D^T & M_R
\end{array}\right). \label{massm}
\end{equation}
Typically, Dirac masses $M_D$ are much smaller than Majorana masses $M_R$ i.e. $M_D\ll M_R$, e.g. $M_D\sim10^{-3}\,\textrm{GeV}$, $M_R\sim10^3\,\textrm{GeV}$. Hence light neutrinos $\nu_{1,2,3}$ obtain masses $M_{\nu_{1,2,3}}$ of the order of $1\,\textrm{eV}$ via type I see-saw  mechanism. Unitary matrix $U$ which enters Takagi decomposition $M_{\nu}=U^T\mathrm{diag}(M_{\nu_1},M_{\nu_2},M_{\nu_3},M_{N_1},M_{N_2},M_{N_3})U^\dag$ is of the following form:
\begin{eqnarray}\label{V}
U\approx\left(
\begin{array}{cc}
%K_{LA}^T\approx
1 
& 
%K_{LB}^T\approx
0\\
%K_{RA}^\dag\approx
0 & K_{R}^\dag
\end{array}\right),
\end{eqnarray}
where, $K_{R}$ is an unitary $3\times3$ matrix defined by $M_R=K_{R}^T\mathrm{diag}(M_{N_1},M_{N_2},M_{N_3})K_{R}$, $M_{N_a}>0$. 
For simplicity, we assume no light-heavy neutrino mixings (they are negligible or very small \cite{seesaw}). 
Such choice of $U$ means that $W_2^\pm$ does not couple to light neutrinos $\nu_a$, and heavy neutrinos $N_a$ do not couple to $W_1^{\pm}$. Exact neutrino mixing matrix $U$ can also be considered, which include non-zero off-diagonal light-heavy matrix elements in \eqref{V} \cite{Chen:2013foz}. 

$K_R$ matrix enters directly heavy neutrinos - $W_2$ interactions, which can be cast in the following form \cite{Gluza:1993gf}:
\begin{eqnarray}
\mathcal{L}&\supset&\frac{g_L}{\sqrt{2}}\overline{N}_a\gamma^{\mu}P_R(K_{R})_{aj}l_jW_{2\mu}^{+}
%\cos\xi
+\mathrm{h.c.}\label{lagr1}
%\\
%&&-\frac{g_L}{\sqrt{2}}\overline{N}_a\gamma^{\mu}P_R(K_{R})_{aj}l_jW_{1\mu}^{+}\sin\xi+\mathrm{h.c}.
%
%\label{lagr3}\\
%
%&&+\frac{g_L}{\sqrt{2}}\overline{l}_iP_-\gamma^{\mu}
%%\underbrace{
%(K_{R})_{bi}^*
%%}_{(K_{R}^\dag)_{ib}}
%N_bW_{2\mu}^{-}\,.\label{lagr2}
\end{eqnarray}
%We set $\xi=0$ hence the second term is not relevant for the discussed setup. 

In general elements of the $K_R$ matrix can be complex. In a CP-conserving case, 
CP parities of heavy neutrinos are purely imaginary \cite{Bilenky:1987ty,cpph} and, in fact, 
%CP parities of heavy neutrinos 
they can be connected with elements of the $K_R$ matrix.
In practice, if CP parities of all three heavy neutrinos are the same, 
$\eta_{CP}(N_1)=\eta_{CP}(N_2)=\eta_{CP}(N_3)=+i$, then elements of the $K_R$ matrix can all be made real. If, for instance,  $\eta_{CP}(N_1)=\eta_{CP}(N_2)=-\eta_{CP}(N_3)=+i$ then ${K_R}_{i3}$ element is complex. 
%For general discussions including also CP-violating cases, see \cite{cpph}.
Choosing different scenarios have  far-reaching consequences in phenomenological studies.
Let us consider processes where heavy neutrinos propagate as virtual states, then their contributions to the amplitudes must be summed over. In general, 
constructive or destructive interferences between heavy neutrinos can appear. For instance, in the neutrinoless double beta decay $(\beta \beta)_{0\nu}$ process, or its inverse collider version process $e^-e^- \to W^-W^-$, 
amplitudes include squared matrix elements ${(K_R)}^2_{1a}$.   
If all heavy neutrinos have the same CP parities, then elements of the $K_R$ matrix can be made all real, and all heavy neutrinos contribute constructively into the amplitudes, otherwise  destructive interferences can appear. Such scenarios have been considered in full details in phenomenological analyses in \cite{mycp}. 
It has been shown there that cancellations among contributions to the amplitude
from heavy neutrinos with opposite CP parities can appear.  In this way, low energy $(\beta \beta)_{0\nu}$ constraints  can be avoided and  for instance the collider signal $e^-e^- \to W^-W^-$ can be substantial. 
We will see in the next Section that CP phases of heavy neutrinos play a crucial role also in a case of SS and OS $pp \to lljj$ signals.

\section{Cross sections}

We shall show that interference effects, CP phases of heavy neutrinos and their mass splittings are relevant for the prediction of the $pp\to lljj$ cross section. 
%
%To expose interference effects in a clear way, the following three different setups will be discussed: (A) neutrinos have %degenerate masses, (B) there is only small mass splitting and (C) only one of them is lighter than $W_2$. The %numerical analysis has been done with the help of \texttt{MadGraph5} (v2.2.2). 
%
%
To expose interference effects in a clear way, the following three different setups will be discussed: (A) neutrinos have degenerate masses, (B) one neutrino is lighter than $W_2$, (C) two neutrinos are lighter than $W_2$,  and, (D) finally, there is only small mass splitting among neutrinos. The numerical analysis has been done with the help of \textsc{MadGraph5} (v2.2.2) \cite{Alwall:2011uj} and with our implementation of the MLRSM in \textsc{FeynRules} (v2.0.31) \cite{Christensen:2008py,Degrande:2011ua}.

To simplify notation we shall denote cross-sections for the process $pp\to l_i^{\pm}l_j^{\mp}jj$ by $\sigma_{l_il_j}^{\pm\mp}$ etc. For reference points it is assumed, as in CMS \cite{Khachatryan:2014dka} analysis, that $M_N=M_{W_2}/2$ with diagonal and real $K_R$ mixing matrix in (\ref{lagr1}),
%(\ref{lagr2}), 
which for 
$\sqrt{s}=8\,\mathrm{TeV}$ and  
%$v_R=4.77
%\,(5.21)
%\,\mathrm{TeV}$, 
%what gives 
$M_{W_2}=2.2
%\,(2.4)
\,\mathrm{TeV}$, gives:
\beqa\label{ppW2}
%\sigma(\texttt{\small p p > w2$\pm$})
\sigma(p p \to  W_2^\pm)
=\left\lbrace
\begin{array}{l}
71.16
%\,(36.32)
\,\mathrm{fb},\\
21.09
%\,(10.5)
\,\mathrm{fb},
\end{array}
\right.
\eeqa
what agrees with recent estimations on $pp\to W_2\to jj$ cross section \cite{Aad:2014aqa}. For chosen value of $v_R$ and diagonal matrix $K_R$ relevant branching ratios are:
\beqa
%\label{BRdeg}
\mathrm{BR}(W_2^\pm\to e^\pm N_a)=0.058,\label{BRWdeg}
\eeqa
\beqa
\mathrm{BR}(N_a\to e^\pm jj)=0.35\label{BRNdeg}
\eeqa
when all heavy neutrinos have the same mass $M_N=M_{W_2}/2$, and
\beqa
\mathrm{BR}(W_2^\pm\to e^\pm N_1) =0.066
\eeqa
when only $M_{N_1}=M_{W_2}/2$ while $M_{N_{2,3}}$ are heavier than $W_2$. 

\subsection{Degenerate masses of heavy neutrinos
%$M_{N_J}<M_{W_2}$
}

First, let us examine the following mass pattern 
%with degenerate $M_{N_J}=M_N$. 
%First, we shall consider a setup 
in which all heavy neutrinos are degenerate and lighter than $W_2$: 
\begin{equation}
M_N:=M_{N_a}=M_{W_2}/2. 
%< M_{W_2} 
\end{equation}
In this setup, and also  for small mass differences between heavy neutrinos, the narrow width approximation (NWA) will not work because of the interference effects.  
%[E. Fuchs, Cargese2014 talk]
%We want to show what can be the possible source of discrepancy between $\sigma^{\pm\mp}$ and $\sigma^{\pm\pm}$. We start from standard and general  parametrization for $K_R$, analogous as for unitary $3\times 3$ light neutrino mixing matrix, its general form is given in (\ref{3times3}). It can be done so, as we assume the complete $6\times 6$ mixing matrix in diagonal block form, see (\ref{V}). To this end, we shall 

Let us take $K_{R}$ in the following form (which is in fact a product of real, orthogonal transformation and diagonal phase matrix) 
%\cite{Bilenky:1987ty,cpph,mycp}
%
%(which is 
%explicit embedding of $U(1)\times U(2)\subset U(3)$ and 
%the most general form of mixing matrix in the case when the heaviest neutrino $N_3$ interact only with $\tau$)
\beq\label{KRB22}
K_{R}=
%e^{i\alpha_{11}}
\left(
\begin{array}{ccc}
\cos\theta_{12}&
%e^{i\gamma}
\sin\theta_{12}&0\\
-e^{i\phi_2}\sin\theta_{12}&e^{i
%(
%\gamma+
\phi_2
%)
}\cos\theta_{12}&0\\
0&0&1
\end{array}
\right).
\eeq 
%$\gamma=\alpha_{12}-\alpha_{11}$ \textbf{does not appear in the cross-section hence can be set to 0 ?}, while  
This is a simplified version of a complete unitary rotation matrix \cite{Maki:1962mu}.  
%\cite{Bilenky:1987ty,cpph,mycp}, which is given in the Appendix, (\ref{3times3}).

In this way, we assume mixings between two lepton flavors only. 
Phase 
$\phi_2
%=\alpha_{23}-\alpha_{13}
$ is connected with CP parity of heavy neutrinos $N_{1,2}$, CP-conserving case is realized when $\phi_2=0, \pm \pi/2, \pm \pi$. 
All phases which do not fulfill the above relations break CP symmetry. In general, in the MLRSM with the mass matrix of the form (\ref{massm}) we have six CP phases; if  $v_L \neq 0$ $(M_L\neq 0)$ then there are 18 CP phases \cite{aq}.

Using this simple form of the matrix $K_R$ we are already able to discuss all relevant effects connected with mixings and CP phases in the considered process.
  
%There are simple consequences of choosing~\eqref{KRB22}. 

First, in the case of degenerate neutrinos, 
$\sigma_{l_il_j}^{\pm\mp}$ with $i=j$ does not depend on mixing angles at all, and is zero for $i\neq j$:
\beqa
\label{deltaij}
%\sigma_{i-j+}&=&\\
\sigma^{\pm\mp}_{l_il_j}&=&\frac{g_L^4}{4}
%\\
%&&\times
\left|\sum_{a}F^{\pm\mp}(s,M_{W_2}^2,M_N^2)(K_{R}^\dag)_{ia}(K_{R})_{aj}\right|^2 \nonumber \\
&=&\delta_{ij}\frac{g_L^4}{4}F^{\pm\mp}(s,M_{W_2}^2,M_N^2)=\delta_{ij}\widehat{\sigma}_{SF}^{\pm\mp},
\eeqa
where the second equality comes from the unitarity of the $K_{R}$, while the third defines $\widehat{\sigma}_{SF}^{\pm\mp}$. $F^{\pm\mp}$ is a function of center of mass energy $\sqrt{s}$ and masses $M_{W_2}$ and $M_N$ (leptons and constituents of jets are treated as massless). 
From now on we will not write down arguments of the $F$ functions. 
%
%Note that the presence of events with $e^+\mu^-$ and $e^-\mu^+$ in the final state identified as coming from $W_2$ and degenerate $N_J$ decays will signal %that $K_{R}$ is not an unitary matrix {\bf ??? na pewno? zobacz uwage po (12)}, what is, in general, possible but needs non-zero off-diagonal parts of \eqref{V}.  

\begin{figure}[h!]
\begin{center}
\includegraphics[scale=0.5]{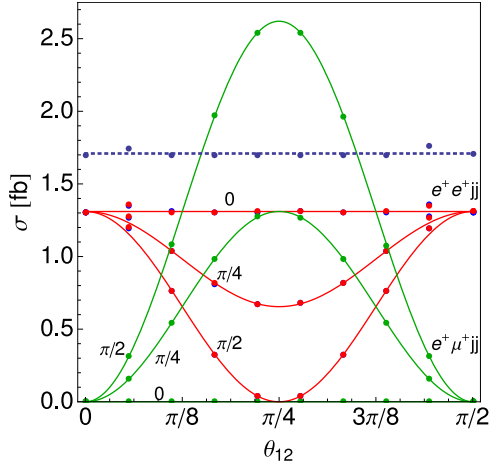}
\caption{Cross section $pp\to ll jj$ for the production of two SS light leptons with two jets $jj$ for $\phi_2=0$, $\pi/4$ and $\pi/2$: $e^+e^+$ (red), $\mu^+\mu^+$ (red, same as for $e^+e^+$) and $e^+\mu^+$ (green). Plots for $\sigma^{--}$ are of the same shape but with $\widehat{\sigma}_{SF}^{++}$ changed by
$\widehat{\sigma}_{SF}^{--}$  in \eqref{sigmaApmpm}. Solid lines show formulas \eqref{sigmaApmpm}, while green and red dots are numerical results obtained in \textsc{madgraph5}. 
%
%Some figure exhibiting interference effects for the degenerate neutrinos scenario. 
%For $l_i^-l_j^+$ there will be no interference and no dependence on mixing parameters while for $l_i^-l_j^-$ and $l_i^+l_j^+$ it will be clearly visible.
The blue dotted line shows corresponding cross section for the OS process $pp\to W_2^{\pm}\to e^+e^-jj$, which is independent of $\theta_{12}$ -- see (\ref{deltaij}).  
}\label{sigmaA}
\end{center}
\end{figure}

On the other hand, for same-sign signature i.e. $l^+l^+$ or $l^-l^-$ the mixing matrix $K_{R}$ does not cancel from the cross section formula: 
\beqa
\sigma_{l_il_j}^{\pm\pm}&=&\frac{g_L^4}{4}\left[F^{\pm\pm}_{1}+(-1)^{\delta_{ij}}F^{\pm\pm}_{2}\right]\nonumber\\
&&\times \left|\sum_{a}(K_{R}^\dag)_{ia}(K_{R}^*)_{aj}\right|^2.
\eeqa
%where $F^{\pm\pm}_{1,2}$ are functions of $s$, $M_{W_2}^2$ and $M_N^2$. 
As a consequence, cross section for 
$l_i^{\pm}l_j^{\pm}$ with $i=j$ is correlated with that for which $i\neq j$ i.e. 
\beqa\label{sigmaApmpm}
\sigma_{l_il_j}^{\pm\pm}&=&\left\lbrace
\begin{array}{lcc}
%\sigma_{SF}|\cos^2\theta_{12}+e^{\pm2i\phi}\sin^2\theta_{12}|^2\\
%&&=
\widehat{\sigma}_{SF}^{\pm\pm}(1-\sin^22\theta_{12}\sin^2\phi_2)&\textrm{for}&i=j,\\[1.5\baselineskip]
%&&\\
%\sigma(l_i^{\pm}l_j^{\pm})&=&
\widehat{\sigma}_{DF}^{\pm\pm}\sin^22\theta_{12}\sin^2\phi_2&\textrm{for}&i\neq j,
\end{array}\right.
\eeqa
where $\widehat{\sigma}_{SF}^{\pm\pm}$ and $\widehat{\sigma}_{DF}^{\pm\pm}$ correspond to 
%`no mixing' and `zero CP phases' scenario
maximal values of cross sections
for same-flavour (SF) and different flavour (DF) cases. For the numerical results see Fig. \ref{sigmaA}. 
The difference in $\widehat{\sigma}_{DF}^{\pm\pm}$ and $\widehat{\sigma}_{SF}^{\pm\pm}$ is related to the standard factor of $(-1)$ appearing in same-flavor Feynman diagrams.  For $\sqrt{s}=8\,\mathrm{TeV}$ and $M_N=1.1\,\mathrm{TeV}$ they read $\widehat{\sigma}_{SF}^{++}=1.31\,\mathrm{fb}$, $\widehat{\sigma}_{SF}^{--}=0.39\,\mathrm{fb}$, $\widehat{\sigma}_{DF}^{++}=2.61\,\mathrm{fb}$, $\widehat{\sigma}_{DF}^{--}=0.78\,\mathrm{fb}$ and $\widehat{\sigma}^{\pm\mp}_{SF}=1.70\,\mathrm{fb}$. $\widehat{\sigma}^{--}$ is about 3.4 times smaller than $\widehat{\sigma}^{++}$ due to asymmetry in production of $W_2^\pm$, see \eqref{ppW2}. 

As one can see from \eqref{sigmaApmpm} and Fig. \ref{sigmaA}, there exists a CP phase for which $(\sigma_{ee}^{++}+\sigma_{ee}^{--})/\sigma_{ee}^{+-}=r$ as suggested by CMS data (\ref{r}). Namely, that relation holds when  $\theta_{12}$ and $\phi_2$ satisfy
\beqa\label{cr}
\sin^22\theta_{12}\sin^2\phi_2=1-\frac{r}{c},
\eeqa
where $c=(\widehat{\sigma}_{SF}^{++}+\widehat{\sigma}_{SF}^{--})/\widehat{\sigma}_{SF}^{+-}\approx
%0.99
1$. As a consequence same-sign different-flavour cross section is $\sigma_{e\mu}^{\pm\pm}=\widehat{\sigma}_{DF}^{\pm\pm}(1-r/c)$. 
%At the same time 
Moreover the total cross section for $pp\to eejj$ is then
\beq\label{sigeetot}
\sigma_{ee}^{(\mathrm{tot})}=\widehat{\sigma}_{SF}^{\pm\mp}[1+r-r\sin^22\theta_{12}\sin^2\phi_2].
\eeq 
%To reconcile CMS data and suppress \eqref{sigeetot} by a factor of $\gamma$ mentioned in the Introduction,  %the following relation has to be fullfiled:
%
%In consequence of \eqref{cr}, 
One can check that the total cross section $\sigma_{ee}^{(\mathrm{tot})}$ is suppressed by a factor
\beq
\gamma=\frac{1+r}{1+c}\approx0.54
%\gamma=1-\frac{r}{1+r}\sin^2\theta_{12}\sin^2\phi. 
\eeq
with respect to $\theta_{12}=\phi_2=0$ case ($\sigma_{ee}^{(\mathrm{tot},0)}$). Our numerical calculations yield $\sigma_{ee}^{(\mathrm{tot},0)}=3.41\,\mathrm{fb}$.
%when $r=1/14$
%for $\phi=0$ it is not possible to satisfy that relation while for $CP$-phase  $\phi=\pi/2$ the angle $\theta_{12}$ is fixed by 
%\beq
%\sin^2\theta_{12}=\frac{1}{r}(1+r)(1-\gamma).
%\eeq
%the maximal suppresion in this scenario is:
%\beq
%\gamma_{max}=1-\frac{r}{1+r}.
%\eeq
Hence 
%it is not possible to find $\theta_{12}$ and $\phi$ such that simultaneously $r=1/14$ and $\gamma=0.23$, in %the case of degenerate masses of $N_a$.
%
when $\theta_{12}$ and $\phi_2$ are chosen such that \eqref{cr} is satisfied then
\beq\label{stot0ee}
\sigma_{ee}^{(\mathrm{tot})}=\gamma\sigma_{ee}^{(\mathrm{tot,0})}=1.84\,\mathrm{fb}
\eeq 
what is about 81\% of the excess reported by the CMS (point B on Fig.~\ref{CMS}). Moreover, in consequence of \eqref{sigmaApmpm}  total cross section for production of two muons with two jets is the same as for electrons:  $\sigma_{ee}^{(\mathrm{tot})}=\sigma_{\mu\mu}^{(\mathrm{tot})}$. Hence the discussed scenario would also result in excess in $\sigma(pp\to \mu\mu jj)$.  That is in contradiction with the CMS data related to $pp\to \mu\mu jj$ \cite{Khachatryan:2014dka}.

\subsection{$M_{N_1}<M_{W_2}<M_{N_{2,3}}$}

In this case only $N_1$ can be on-shell. 
%Mass pattern $M_{N_{1,2,3}}<M_{W_2}$ has been already , $M_{N_{1,3}}<M_{W_2}<M_{N_2}$ and $M_{N_{1,3}}<M_{N_2}<M_{W_2}$ have been discussed in the previous subsections while in the case $M_{W_2}<M_{N_{1,2,3}}$ heavy neutrinos are too heavy to be on-shell and their contributions to the cross section are suppressed. 
%{\bf How much they are suppressed?}  
%
%We assume that only one heavy neutrino, $N_1$, is lighter than $W_2$. More precisely, 
We choose $M_{N_1}=1.1\,\mathrm{TeV}=M_{W_2}/2$; % and $M_{W_2}=2.2\,\mathrm{TeV}$. 
 the remaining two neutrinos are much heavier, $M_{N_{2,3}}=10\,\mathrm{TeV}$. %The role of mixing parameters will be discussed later on.

Here one can use narrow width approximation (NWA) to estimate cross-section for $pp\to l_1l_2jj$ going through on-shell $W_2$, which decays to $l_i$ and on-shell $N_1$ and the latter decays to $l_jjj$: 
% can be written in the following form:
\beqa\label{NWA}
%\sigma(pp\to l_il_jjj)
\sigma_{l_il_j}
&=&\sigma(pp\to W_2)\mathrm{BR}(W_2\to l_iN_1)\nonumber\\
&&\times \mathrm{BR}(N_1\to l_jjj).
\eeqa

Since quarks and leptons masses are much smaller than the $N_1$ mass, 3-body decay of $N_1$ mediated by off-shell $W_2$ can be treated analogously to well-known muon decay in the Fermi theory. One can check that 
\beqa\label{Nwidth}
\Gamma(N_a\to l_i^- q_{\alpha}\overline{q}_{\beta})&=&\frac{g_L^4}{2048\pi^3}|(K_{R})^{*}_{ai}|^2|(U_{CKM}^R)_{\alpha\beta}|^2\nonumber\\
&&\times M_{N_a}F\left(x_a\right),
\eeqa
where $x_a=M_{N_a}^2/M_{W_2}^2$ while the function 
\beqa
F(x)=\frac{12}{x}\left[1-\frac{x}{2}-\frac{x^2}{6}+\frac{1-x}{x}\ln(1-x)\right]
\eeqa
encompasses full tree-level contribution from the $W_2$ propagator \cite{Ferroglia:2013dga}. The presence of such a factor makes $N_a$ decay width really sensitive to the ratio $x_a=M_{N_a}^2/M_{W_2}^2$, e.g. for fixed $M_{W_2}$ it can be enhanced by a factor of 
%\footnote{$\frac{\Gamma|_{x=1}}{\Gamma|_{x=1/4}}=\frac{M_{N_a}|_{x=1}F(1)}{M_{N_a}|_{x=1}F(1/4)}=\frac{(M_{N_a}|_{x=1}/M_{W_2})F(1)}{(M_{N_a}|_{x=1}/%M_{W_2})F(1/4)}=\frac{2}{F(1/4)}\approx27.107$} 
$\sim 27$ when $M_{N_a}\approx M_{W_2}$ with respect to the scenario $M_{N_a}\approx M_{W_2}/2$. 
%
%{\bf Vasquez did not take that $F(x)$ function into account.} 
%
%{\bf Analogously,} 
%\beqa
%\Gamma(N_J\to l_i^+ \overline{q}_aq_b)&=&\frac{g_L^4}{2048\pi^3}|(K_{R})^{*}_{Ji}|^2|(U_{CKM}^R)_{ab}|^2 \nonumber\\
%&&\times M_{N_J}F\left(x_J\right) (?)
%\eeqa
%

Summing over all possible final states and taking into account the unitarity of $K_{R}$ and $U_{CKM}^R$ one obtains the total decay width of $N_a$ 
\beqa
\Gamma(N_a)&=&\sum_{i,\alpha\beta}\left[\Gamma(N_a\to l_i^- q_{\alpha}\overline{q}_{\beta})+\Gamma(N_a\to l_i^+ \overline{q}_{\alpha}q_{\beta})\right]\nonumber\\
&=&\frac{9g_L^4}{1024\pi^3}M_{N_a}F\left(x_a\right)
\eeqa
Hence the $\mathrm{BR}$s under consideration are 
\beqa\label{BR_N}
\mathrm{BR}(N_a\to l_i^-q_{\alpha}\overline{q}_{\beta})&=&\frac{1}{6}|(K_{R})^{*}_{ai}|^2|(U_{CKM}^R)_{\alpha\beta}|^2,
\eeqa
\beqa\label{BR_Nb}
\mathrm{BR}(N_a\to l_i^+\overline{q}_{\alpha}q_{\beta})&=&\mathrm{BR}(N_a\to l_i^-q_{\alpha}\overline{q}_{\beta}).
%\,\,\textbf{(Check it.)}
\eeqa
%It is worthwhile to note that this branching ratio, which is crucial for $pp\to eejj$, can be tuned not only by modifying $K_{R}$ but also by $U_{CKM}^R$ because  %summing over light quarks only ($\alpha,\beta=1,2$), to approximate decay to two jets, will not cancel dependence on $U_{CKM}^R$: $\sum_{\alpha,\beta=1,2}|%(U_{CKM}^R)_{\alpha\beta}|^2\neq1$.   
%{\bf cos mi to smierdzi...}

Using assumed masses, we have  scanned over $\theta_{12}\in\left<0,\pi/2\right>$ to verify dependence of $\sigma_{l_il_j}
%(pp\to W_2^\pm
%\to xN_1
%\to l_il_jjj)
$  on that angle. The CP phase $\phi_2$ was  set to $\pi/2$ (CP-conserving case), i.e. 
\beq
K_{R}=\left(
\begin{array}{ccc}
\cos\theta_{12}&\sin\theta_{12}&0\\
-i\sin\theta_{12}&i\cos\theta_{12}&0\\
0&0&1
\end{array}
\right).
\eeq

\begin{figure}[h!]
\vspace{2mm}
\begin{center}
\includegraphics[scale=0.5]{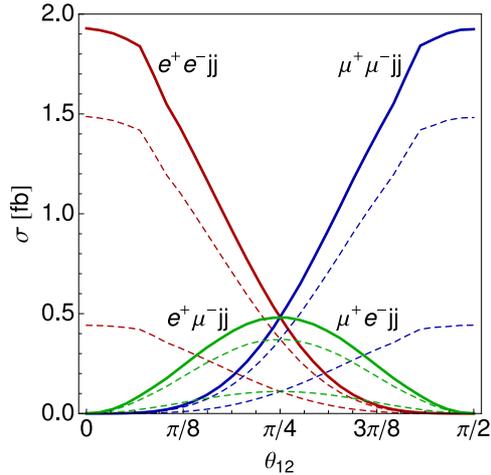}
\caption{Cross section $\sigma$ for the production of two opposite-sign light leptons $l_i=e,\mu$ with two jets $jj$ in the process $pp\to W_2^\pm
%\to x^\pm N_1
\to l_i^+ l_j^- jj$ with $\sqrt{s}=8$ TeV for $M_{N_1}=M_{W_2}/2$, $M_{N_{2,3}}>M_{W_2}$. The dashed lines display contributions 
%to processes coming 
from intermediate channels $W_2^\pm\to e^\pm N_1$ and $W_2^\pm\to \mu^\pm N_1$. 
%, where $x$ is $e$ or $\mu$. 
Solid lines correspond to sum over possible channels.
}
\label{OS-th12}
\end{center}
\end{figure}
\begin{figure}[h!]
\vspace{2mm}
\begin{center}
\includegraphics[scale=0.5]{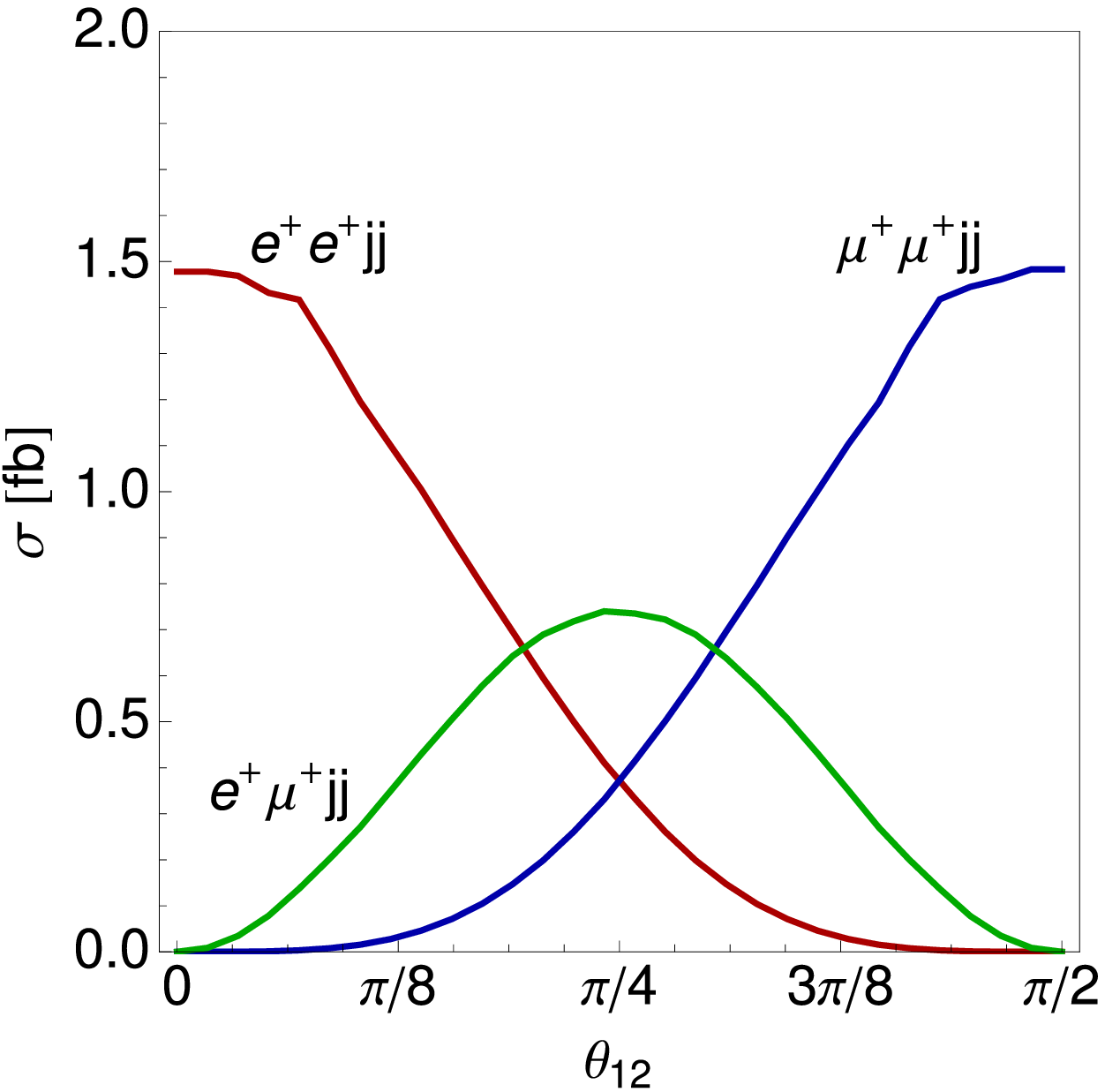}
\caption{
Cross section $\sigma$ for the production of two same-sign light leptons $l_i^+=e^+,\mu^+$ with two jets $jj$ in the process $pp\to W_2^+
%\to x^+ N_1
\to l_i^+ l_j^+ jj$ with $\sqrt{s}=8$ TeV for $M_{N_1}=M_{W_2}/2$, $M_{N_{2,3}}>M_{W_2}$. 
%The dashed lines display contributions to processes coming from intermediate channels $W_2^+\to e^+ N_1$ and $W_2^+\to \mu^+ N_1$. It is the same line for both $e^+$ and $\mu^+$. 
%{\bf Rethink the issue of interference between diagrams with different intermediate channels.}
%The dashed lines displays contributions to processes coming from intermediate channels $W_2^\pm\to x^\pm N_1$, where $x$ is $e$ or $\mu$. Solid lines correspond to sum over possible channels.
}
\label{SSp-th12}
\end{center}
\end{figure}
\begin{figure}[h!]
\vspace{2mm}
\begin{center}
\includegraphics[scale=0.5]{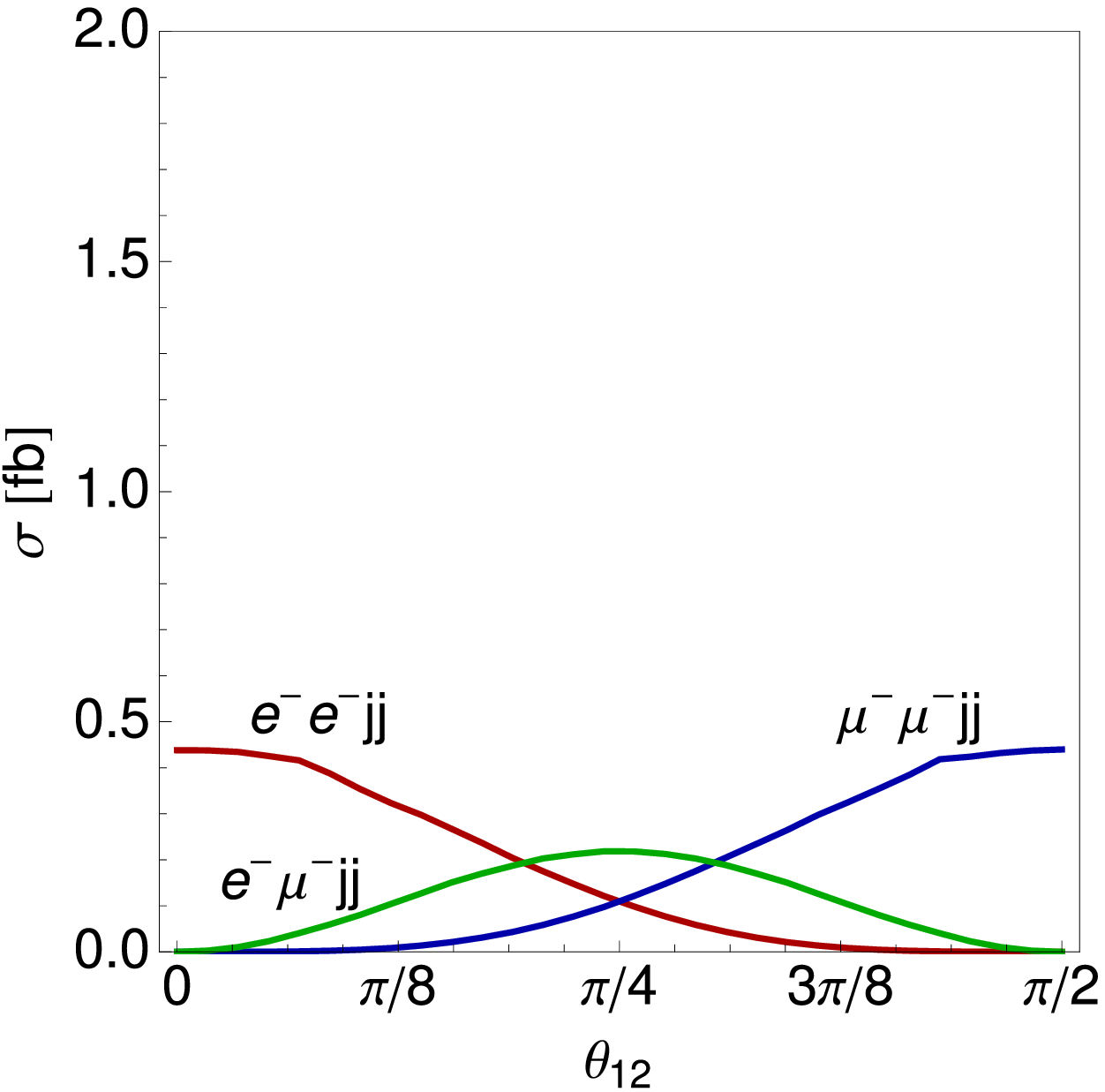}
\caption{
Cross section $\sigma$ for the production of two same-sign light leptons $l_i^-=e^-,\mu^-$ with two jets $jj$ in the process $pp\to W_2^-
%\to x^- N_1
\to l_i^- l_j^- jj$ with $\sqrt{s}=8$ TeV for $M_{N_1}=M_{W_2}/2$, $M_{N_{2,3}}>M_{W_2}$. 
%The dashed lines display contributions to processes coming from intermediate channel $W_2^-\to e^- N_1$ and $W_2^-\to \mu^- N_1$. It is the same for both $e^-$ and $\mu^-$.
}
\label{SSm-th12}
\end{center}
\end{figure}

\begin{figure}[h!]
\vspace{2mm}
\begin{center}
\includegraphics[scale=0.5]{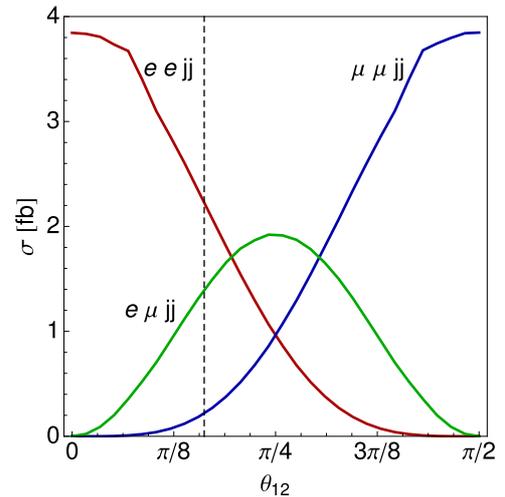}
\caption{
The total cross section $\sigma$ for the production of two light leptons $l_i=e,\mu$ with two jets $jj$ in the process $pp\to W_2
\to l_il_jjj$ with $\sqrt{s}=8$ TeV for $M_{N_1}=M_{W_2}/2$, $M_{N_{2,3}}>M_{W_2}$. The vertical dashed line displays value of $\theta_{12}$ for which $\sigma_{ee}^{(\mathrm{tot})}$ (red solid line) matches CMS excess value (point B on Fig.~\ref{CMS}). 
}
\label{tot-th12}
\end{center}
\end{figure}

The obtained dependences are shown on the Figs. \ref{OS-th12}, \ref{SSp-th12} and \ref{SSm-th12}. On these plots, we present contributions to the total  cross section $\sigma_{l_il_j}$ from subprocesses with different charges and flavors of leptons in the final state. The scale on the vertical axes is the same for all these plots to clearly show relative values of individual cross sections. The total cross section itself is shown on Fig.~\ref{tot-th12}. 

Let us first note that there is no interference between different contributions to  $pp\to l_i^+l_j^-jj$, see Fig.~\ref{OS-th12}, because the corresponding initial states (at the parton level) are different. 
Secondly, due to their large masses $N_{2,3}$ are decoupled and effectively only contributions from Feynman diagrams containing $N_1$ are relevant. In this case NWA can be used to understand qualitative dependence on the mixing angle $\theta_{12}$. Namely, using \eqref{NWA} one obtains $\sigma_{ee}\sim\cos^4\theta_{12}$, $\sigma_{\mu\mu}\sim\sin^4\theta_{12}$ and for different-flavor signature $\sigma_{e\mu}\sim\sin^22\theta_{12}$, cf. Figs.~\ref{OS-th12}, \ref{SSp-th12} and \ref{SSm-th12}. Thirdly, one can check that, due to decoupling of $N_{2,3}$, in this setup CP phases do not influence cross sections because the interference between diagrams with different $N_a$ is suppressed by large mass of $N_{2,3}$. It is worthwhile to note that here, as in Sec. A, the difference between maximal value of $\sigma_{ee}^{\pm\pm}$ and $\sigma_{e\mu}^{\pm\pm}$, see Figs.~\ref{SSp-th12} and \ref{SSm-th12}, comes from the standard factor of $(-1)$ appearing in same-flavor Feynman diagrams. 
Finally, our numerical analysis shows that in this scenario $\sigma_{ee}^{(\mathrm{tot},0)}=3.89\,\mathrm{fb}$ hence to address CMS excess in $\sigma_{ee}^{(\mathrm{tot})}$ one has to adjust $\theta_{12}$ to $0.51$. 
%Obviously, other choice of CP phase $\phi$ would result in different mixing matrix elements.  
At the same time $\sigma_{\mu\mu}^{(\mathrm{tot})}=0.21\,\mathrm{fb}$, see Fig.~\ref{tot-th12}, so there is no excess in the $\mu\mu jj$ what is in accordance with CMS data \cite{Khachatryan:2014dka,Khachatryan:2015gha}. However as one can check, cf. Figs.~\ref{SSp-th12} and \ref{SSm-th12}, sum of same-sign signature cross sections i.e. $\sigma_{ee}^{++}+\sigma_{ee}^{--}$ is nearly equal to $\sigma_{ee}^{+-}$ for all values of mixing angle $\theta_{12}$. As a consequence, in this setup $r\approx1$ and one cannot address \eqref{r} by adjusting $\theta_{12}$.  

%{\bf  Remove dashed lines, but keep that version of the plots in the folder of .tex file.} 
%{\bf [Co z nich wynika? Co daja te trzy wykresy osobno i razem? Czy istnieja katy $\theta_{12}$ i $\phi$ takie, ze $r$ mozna wytlumaczyc oraz brak $\mu\mu$?]}

\subsection{$M_{N_{1,3}}<M_{W_2}<M_{N_{2}}$}

However, it turns out that one can arrange parameters of the models such that  all above-mentioned experimental constraints are fulfilled. Namely, let us now consider the following mass pattern: 
\beq
M_{N_{1,3}}=0.925\,\mathrm{TeV},\quad M_{N_2}=10\,\mathrm{TeV}
\eeq
and mixing matrix of the form:
\beq\label{KRB13}
K_{R}=
\left(
\begin{array}{ccc}
\cos\theta_{13}&0&\sin\theta_{13}\\
0&1&0\\
-e^{i\phi_3}\sin\theta_{13}&0&e^{i\phi_3}\cos\theta_{13}
\end{array}
\right).
\eeq 
One expects that here $\mu\mu jj$ signal should be suppressed due to the large mass of $N_2$. In fact, it is confirmed by numerical computations:  $\sigma_{\mu\mu}^{(\mathrm{tot},0)}\approx0\,\mathrm{fb}$ while $\sigma_{ee}^{(\mathrm{tot},0)}=4.21\,\mathrm{fb}$. Because in this scenario $N_1$ and $N_3$ are degenerate in masses, one also gets: $\sigma_{\tau\tau}^{(\mathrm{tot})}=\sigma_{ee}^{(\mathrm{tot})}$. Let us note that here $\mathrm{BR}(W_2^\pm\to e^{\pm}N_{1,3})=0.071$ due to $x_{1,3}=M_{N_{1,3}}^2/M_{W_2}^2\approx0.18$ and $x_3=M_{N_2}^2/M_{W_2}^2>1$, see Appendix. That enhancement of $\mathrm{BR}$ with respect to \eqref{BRWdeg} compensates deficit in \eqref{stot0ee}.  As previously, analysis of contributions from heavy neutrinos $N_{1,3}$ gives $\sigma^{\pm\mp}_{l_il_j}=\delta_{ij}\widehat{\sigma}_{SF}^{\pm\mp}$ and: 
\beqa\label{sigmaA1pmpm}
\sigma_{l_il_j}^{\pm\pm}&=&\left\lbrace
\begin{array}{lcc}
\widehat{\sigma}_{SF}^{\pm\pm}(1-\sin^22\theta_{13}\sin^2\phi_3)&\textrm{for}&i=j,\\[1.5\baselineskip]
\widehat{\sigma}_{DF}^{\pm\pm}\sin^22\theta_{13}\sin^2\phi_3&\textrm{for}&i\neq j,
\end{array}\right.
\eeqa
where $i,j\in\{1,3\}$. Now the maximal values of cross sections are: $\widehat{\sigma}_{SF}^{\pm\mp}=2.14\,\mathrm{fb}$, $\widehat{\sigma}_{SF}^{++}=1.63\,\mathrm{fb}$, $\widehat{\sigma}_{SF}^{++}=0.48\,\mathrm{fb}$  and $\widehat{\sigma}_{DF}^{++}=3.27\,\mathrm{fb}$, $\widehat{\sigma}_{DF}^{--}=0.96\,\mathrm{fb}$. Moreover, 
\beq\label{th13eq}
\sin^22\theta_{13}\sin^2\phi_3=1-\frac{r}{c}
\eeq
has to be satisfied in order to ensure $r=1/14$. As previously, $c=(\widehat{\sigma}_{SF}^{++}+\widehat{\sigma}_{SF}^{--})/\widehat{\sigma}_{SF}^{+-}\approx1$ and   $\gamma\approx0.54$ what gives $\sigma_{ee}^{(\mathrm{tot})}=\gamma\sigma_{ee}^{(\mathrm{tot},0)}=2.27\,\mathrm{fb}$. It is precisely the value of $\sigma(pp\to eejj) $ reported by the CMS (point B on Fig.~\ref{CMS}). In this way both the lepton flavor and charge independent results as well as OS (electron) dominance over SS (muon) signals can be recovered. It happens for $\theta_{13}$ and $\phi_{13}$ values which satisfies Eq.~\eqref{th13eq}.

%
%But it is not possible to reduce the total cross section for $pp\to eejj$ to the level of $2\,\mathrm{fb}$, because its minimal possible value is $\sigma_{ee}^{+-}$/it %is not possible address value of $\sigma$ related to CMS excess. 
%{\bf skomplikowane relacje, nie kumam, czy mozesz zrobic tak aby CP faza dawala CP-conserving case (czyli np phi=pi/2), dla jakiego theta12 bedzie odpowiednie r?}

Let us remark that naive usage of NWA would not capture dependence on CP phases $\phi_{2,3}$ at all, neither interference between diagrams with different $N_a$ correctly, what will result in wrong $\theta_{12,13}$ dependence, nor contributions from 
%$u$-channel (
diagram with crossed lepton lines in the case of same-flavour  signature.
%
%This is a very important issue in data analysis. 
%
This should be kept in mind when confronting refined models with data. 
%
%{\bf [Czy to mozna jakos latwo wytlumaczyc dlaczego jest problem?]}

\subsection{Dependence on heavy neutrino mass splitting $\Delta M=M_{N_2}-M_{N_1}$}

Here we want to show some general dependence of the cross section on mass difference between $N_2$ and $N_1$. For simplicity it is assumed that mass of the first and  third heavy neutrino are fixed to $1\,\mathrm{TeV}$.  

\begin{figure}[h!]
\begin{center}
\includegraphics[scale=0.5]{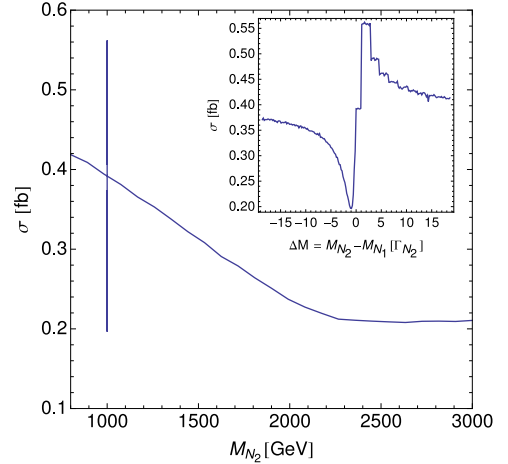}
\caption{Cross section $pp \to e^+e^+jj$ with $\theta_{12}=\phi=\pi/4$.  Subplot on this figure exhibits interference effects for small splitting in masses of heavy   neutrinos. Note that on the subplot mass difference $\Delta M=M_{N_2}-M_{N_1}$ is expressed in terms of multiplicities of $N_2$ decay width $\Gamma_{N_2}\approx0.53\times 10^{-3}\,\mathrm{GeV}$. 
%For $l_i^-l_j^+$ there is no interference and no dependence on mixing parameters while for $l_i^-l_j^-$ and $l_i^+l_j^+$ it is clearly visible. 
%The vertical line at $M_{N_2}=\,\mathrm{TeV}$ is a 'squeezed' version of the curve displayed on the subplot. 
%Zooming region around $M_{N_2}=1\,\mathrm{TeV}$, showed on the subplot, reveals interference-like shape of the the curve. {\bf Is the behaviour of $\sigma$ consistent with NWA? Other values of $\phi$ and $e^-e^-$, $e^+e^-$ cases.}
} 
\label{N2-zoom}
\end{center}
\end{figure}

Let us note first that  $\sigma$ decreases when $M_{N_2}\to M_{W_2}$, see Fig. \ref{N2-zoom}. 
%
%The decreasing of $\sigma$ when $M_{N_2}$ grows
It is a consequence of decreasing branching ratio, see (\ref{wnl}) in the Appendix. 
%
%\begin{equation}
%\mathrm{BR}(W_2^+\to l_i^+ N_J)=|(K_{R}^\dag)_{iJ}|^2\frac{F_W(x_J)}{18+\sum_K F_W(x_K)}.
%\end{equation}
%when $x_2=M_{N_2}^2/M_{W_2}^2$ grows (for functions definition, see (\ref{wn1}).
This effect is substantial; the cross section can be suppressed by a factor of 2 for considered masses. 

%{\bf [Check it numerically. On Fig. \ref{N2-zoom} $\sigma$ seems to decrease linearly with $M_{N_2}$ while the actual plot of $\mathrm{BR}$ as a function of $M_{N_2}$ is non-linear. It may be related to (-1) factor and/or more complicated kinematics in case of same-sign flavour signature. Maybe NWA cannot be applied here even far from interference region. Compare it with shapes of $\sigma^{++}_{e\mu}$ and $\sigma^{+-}_{ee}$].}
%
When $M_{N_2}>M_{W_2}$ then the decay $W_2\to lN_2$ is kinematically forbidden. It means that $N_2$ cannot be on-shell hence the contribution from such a diagram is very small because it is not enhanced by the $N_2$-resonance, cross section starts to be flat in Fig.~\ref{N2-zoom}.

The second effect worth mentioning, is constructive or destructive interference between diagrams with $N_1$ and $N_2$ when $M_{N_2}$ goes across $M_{N_{1,3}}$. Due to very small width of $N_a$, $\Gamma(N_a)\sim10^{-3}\,\mathrm{GeV}$, the interference effect is visible only in the `very degenerate' case i.e when mass difference $|M_{N_{1,3}}-M_{N_2}|$ between heavy neutrinos is smaller than about $0.005\,\mathrm{GeV}$.  Let us stress that due to these interference effects cross section $\sigma$ can be suppressed by an additional  factor of $0.5$ or increased by $1.5$, see Fig. \ref{N2-zoom}.  Hence, very small mass splitting between heavy neutrinos can be a source of additional suppression/enhancement of the discussed cross section. However, as a width is very small, it might be difficult to discover such effects experimentally (energy resolution).

\section{Summary and Outlook}
%Motivated by  the recent data from the CMS  showing a small excess in the cross section for $pp \to eejj$, 
We have revisited production of two light charged leptons and two jets in $pp$ collision in the context of the genuine MLRSM with $g_L=g_R$. Taking into account details on the neutrino mass matrix parameters, interesting conclusions can be derived. Recent CMS data showed that:
\begin{itemize}
\item[(i)] there is an excess in the total $pp\to eejj$ cross section at about $M_{W_2}\approx2.2\,\mathrm{TeV}$;
\item[(ii)] there is a suppression of 
%the number of 
same-sign electron pairs 
%events 
with respect to opposite-sign pairs
%events 
in $pp\to eejj$ events;  
\item[(iii)] there is a suppression of muon pairs with respect to electron pair in $pp\to lljj$ events.
\end{itemize}

These facts cannot be explained within the MLRSM with $g_L=g_R$, degenerate heavy neutrino mass spectrum, and no neutrino mixings in $K_R$. 

However, we have shown that all the issues (i)--(iii) listed above can be reconciled with the $g_L=g_R$ MLRSM, if non-degenerate heavy neutrino mass spectrum, neutrino mixings in $K_R$ and CP phases are taken into account. 

%, in accordance with experimental data. Namely, experimental evidence of the suppression of the total cross section %when compared to the prediction with degenerate neutrinos and real mixing matrix elements and suppression of the %SS electron signal over OS electron signal, both can be a sign for presence of CP phases in the leptonic sector and %non-trivial mixing parameters in the $K_R$ matrix.  Similarly,  suppression of the muon signals over electron signals is %a hint for non-degenerate mass pattern in the heavy neutrino sector.  

We also conclude that it is worth to undertake more careful analyses of the neutrino sector when exclusion plots are considered, otherwise too strong limits can be inferred from a simplified scenario (in this case assuming real neutrino mixing matrix
elements with degenerate heavy neutrinos). An example is specific, but conclusions which we can derive are more general
as heavy neutrinos are present within many BSM models.
	
In our analyses we kept $M_{W_2}$ fixed at the CMS value 2.2 TeV, however, in the light of leptogenesis \cite{Dhuria:2015hta}, it would be interesting to check if it is possible to reproduce CMS data  with $M_{W_2}$ shifted up to about
%/bigger than 
3 TeV by 
%loosing 
relaxing
$M_{W_2}-M_N$ mass 
relation ($M_N=M_{W_2}/2$) and exploring wide space of heavy neutrino mixing angles, phases and masses (not necessarily of the degenerate nature), similarly as we have made in this work. It will be worthwhile to study that issue when better statistics is available.

As an outlook, we would like to check more carefully contributions from the scalar sector in MLRSM and confront our scenarios which include heavy neutrino mixing parameters and CP phases with other delicate low-energy data as neutrinoless double beta decay, just to mention \cite{Tello:2010am,Chakrabortty:2012pp,Chakrabortty:2012mh,
Rodejohann:2012xd,Bambhaniya:2013wza,Dev:2013vxa}.  

\section*{Acknowledgements}

We would like to thank Frank Deppisch for useful technical remarks on CMS data related to $eejj$ excess, and Marek Gluza and Robert Szafron for interesting remarks. 
Work is supported by the Polish National Science Centre (NCN) under the Grant Agreement No.\linebreak[4] DEC-2013/11/B/ST2/04023 and under postdoctoral grant No. DEC-2012/04/S/ST2/00003.

\section*{Appendix}

We collect here some basic formulas useful in calculations and basic estimations.

\begin{itemize}
%%\item $g_L=0.651906$, $M_{W_2}=gv_R/\sqrt{2}$
%%
%\item Parametrization of $K_{R}$ matrix 
%\cite{
%%Agashe:2014kda
%Rasin:1997pn}: 
%$K_{R}=U_1UU_2$,
%%\beqa\label{KR}
%%K_{R}=U_1UU_2, 
%%\eeqa
%where $U_1=\mathrm{diag}(1,e^{i(\alpha_{23}-\alpha_{13})},e^{i(\alpha_{33}-\alpha_{13})})$ and $U_2=\mathrm{diag}(e^{i\alpha_{11}},e^{i\alpha_{12}},e^{i\alpha_{13}})$ while
%\begin{widetext}
%\begin{equation}
%U=\left(
%\begin{array}{ccc}
%c_{12}c_{13}&s_{12}c_{13}&s_{13}e^{-i\delta}\\
%-s_{12}c_{23}-c_{12}s_{23}s_{13}e^{i\delta}&c_{12}c_{23}-s_{12}s_{23}s_{13}e^{i\delta}&s_{23}c_{13}\\
%s_{12}s_{23}-c_{12}c_{23}s_{13}e^{i\delta}&-c_{12}s_{23}-s_{12}c_{23}s_{13}e^{i\delta}&c_{23}c_{13}
%\end{array}
%\right) \label{3times3}
%\end{equation}
%\end{widetext}
%and $c_{rs}$, $s_{rs}$ stand for $\cos\theta_{rs}$ and $\sin\theta_{rs}$ respectively. 

\item Quark gauge interactions \cite{Gluza:1993gf}:
% conventions as in Duka-Gluza-Zralek
\beqa
\mathcal{L}\supset \frac{g_R}{\sqrt{2}}\overline{q}_{\alpha}\left({U_{CKM}^R}^\dag\right)_{\alpha\beta}P_-\gamma^\mu q_{\beta}W^-_{2\mu}\nonumber\\
+\frac{g_R}{\sqrt{2}}\overline{q}_{\beta}\gamma^\mu P_+\left(U_{CKM}^R\right)_{\beta\alpha}q_{\alpha}W_{2\mu}^+.
\eeqa

\item 2-body decay contributions to BR were calculated in \textsc{FeynRules}. 
We treat leptons $l_i^{\pm}$ as massless
%\footnote{$\theta$ function in the definition of $F_W$ takes care of kinematic constraints for the decays.}
:
\beqa\label{wn1}
\Gamma(W_2^+\to l_i^+ N_a)=\frac{g_L^2M_{W_2}}{96\pi}|(K_{R}^\dag)_{ia}|^2F_W\left(x_a\right),
%\\
%&&\Gamma(W_2^-\to l_i^- N_a)=\Gamma(W_2^+\to l_i^+ N_a).
\eeqa
where $x_a=M_{N_a}^2/M_{W_2}^2$ and $F_W(x)=(2-3x+x^3)\theta(1-x)$. $\theta$ function in the definition of $F_W$ takes care of kinematic constraints for the decays. Because even for top quark the ratio $M_{q_{\alpha}}^2/M_{W_2}^2$ is of the order of $10^{-2}$ one can treat quarks in the final states as massless. Hence, taking into account $F_W(0)=2$ and summing over colors:
\beqa
\Gamma(W_2^+\to q_{\alpha}\overline{q}_{\beta})&=&\frac{g_L^2M_{W_2}}{16\pi}|({U_{CKM}^R}^\dag)_{\alpha\beta}|^2,
%\quad\textbf{(Check if indices $a$, $b$ are OK.)}
%\nonumber
\\
\Gamma(W_2^-\to \overline{q}_{\alpha}q_{\beta})&=&\frac{g_L^2M_{W_2}}{16\pi}|({U_{CKM}^R}^T)_{\alpha\beta}|^2.
%\quad \textbf{(Check that ${}^T$ with \texttt{FeynRules})}
\eeqa
%The sum over all possible colors of quarks is $3$ and not $9$ because $W_2$ is colorless, hence choosing a color of %$q_{\alpha}$ automatically fixes a color of $\overline{q}_{\beta}$. Now, it is possible to find the total width of $W_2^+$:
%
%
%\footnote{3-body and 4-body decays will be important when $M_{N_a}>M_{W_2}$. Then an off-shell heavy neutrino $N_a$ will mediate e.g. 4-body decay $W_2^%+\to l_i^-l_j^{\pm}q_{\alpha}\overline{q}_{\beta}$.}
%
That yields the total width of $W_2^{\pm}$
\beqa
\Gamma(W_2^\pm)
%\sum_{i,a}\Gamma(W_2^+\to l_i^+ N_a)\nonumber\\
%&&+\sum_{\alpha,\beta}\Gamma(W_2^+\to q_{\alpha}\overline{q}_{\beta})+(\geq3\textrm{-body decays})\nonumber\\
%&&
%&\approx&\frac{g_L^2M_{W_2}}{96\pi}\left[\sum_{i,a}|(K_{R}^\dag)_{ia}|^2F_W\left(x_a\right)+18
%2\times 3\times\underbrace{3}_{\textrm{sum over flavors}}
%\right]\nonumber\\
%&=&
=\frac{g_L^2M_{W_2}}{96\pi}\left[\sum_{a}F_W\left(x_a\right)+18\right]
%\Gamma(W_2^-)&=&\ldots,
\eeqa
%where 
%$x_a=M_{N_a}^2/M_{W_2}^2$ and 
%in the last step the unitarity of $K_{R}$ has been used: $\sum_{i}K^\dag_{ia}K_{ai}=1\,\forall_a$ (no summation over $a$).  
and branching ratio for $W_2^\pm\to l_i^\pm N_a$:
\beqa\label{wnl}
\mathrm{BR}(W_2^\pm\to l_i^\pm N_a)
%\nonumber\\
=|(K_{R}^\dag)_{ia}|^2\frac{F_W\left(x_a\right)}{18+\sum_{c}F_W\left(x_c\right)}.
%\mathrm{BR}(W_2^-\to l_i^- N_J)&=&\ldots.
\eeqa
That formula gives very good estimate of branching ratio; e.g. for $x_1=1/4$, $x_{2,3}>1$ and $K_{R}=1$ one gets $\mathrm{BR}(\ldots)/\mathrm{BR}(\ldots)_{\texttt{MadGraph}}\approx0.0657/0.0659\approx0.997$, and similarily for $x_{1,2,3}=1/4$:\linebreak[4] $\mathrm{BR}(\ldots)/\mathrm{BR}(\ldots)_{\texttt{MadGraph}}\approx0.0581/0.0582\approx0.998$.

The branching ratio for decay to quarks is:
\beqa
%&&
\mathrm{BR}(W_2^\pm\to q_{\alpha}\overline{q}_{\beta})&=&|(U_{CKM}^R)_{\beta\alpha}|^2\nonumber\\
&&\times\frac{6}{18+\sum_{c}F_W\left(x_c\right)}.
%\mathrm{BR}(W_2^-\to \overline{q}_aq_b)&=&\ldots. 
\eeqa
%
%{\bf [Maybe some plots of these BR's as a function of $M_{W_2}\in[2\,\mathrm{TeV},5\,\mathrm{TeV}]$ for fixed values of $M_{N_J}$ e.g. $M_{N_{1,2,3}}=1,2,3\,\mathrm{TeV}$  would be useful.]}
%For the completness we recall cross section for the production of $W_2^{\pm}$, see also [Han et al.?]:
%\beqa
%\sigma(pp\to W_2^{\pm})=\ldots
%\eeqa
%3-body decay widths of heavy neutrinos including $K_{R}$ and $U_{CKM}^R$:
%\beqa
%\Gamma(N_J\to l_i^-q_a\overline{q}_b)&=&\ldots,\nonumber\\
%\Gamma(N_J\to l_i^+\overline{q}_aq_b)&=&\ldots.
%\eeqa

\end{itemize}

%\end{widetext}

\end{document}